\documentclass[aps,prb,twocolumn]{revtex4}
\usepackage{graphicx}
\usepackage{epsfig}
\usepackage{amsmath}
\usepackage{amssymb}
\usepackage{amsfonts}%
\newcommand{\D}{{\rm d}} 
\newcommand{\I}{{\rm i}} 
\newcommand{\Tr}{{\rm Tr}} 
\renewcommand{\Re}{{\rm Re}} 
\renewcommand{\Im}{{\rm Im}} 
\begin{document} 
\title{Spin-Hall effect in a disordered 2D electron-system }  
\author{Roberto Raimondi}
\affiliation{Dipartimento di Fisica "E. Amaldi," Universit\`a di Roma Tre, Via della Vasca Navale 84, 00146 Roma, Italy}
\affiliation{INFN, Laboratori Nazionali di Frascati, P.O. Box 13, 00044 Frascati, Italy}
\author{Peter Schwab}
\affiliation{Institut f\"ur Physik, Universit\"at Augsburg, 86135 Augsburg, Germany}
\begin{abstract} 
We calculate the spin-Hall conductivity for a two-dimensional electron gas
within
the self-consistent Born approximation, varying the strength and type of disorder. 
In the weak disorder limit we find both analytically and numerically a vanishing 
spin-Hall conductivity even when we allow a momentum dependent scattering. 
Separating the reactive from the disspative current response,
we find the universal value $\sigma^R_{sH} = e/8 \pi$ for the
reactive response, which cancels however with the dissipative part  $\sigma^D_{sH} = -e/8 \pi$. 
\end{abstract} 
\pacs{PACS numbers: } 
 
\date{\today} 
\maketitle 
 
Spin-orbit coupling in two dimensional electron systems allows a number of unconventional
transport phenomena,
since charge current and the spin degrees of freedom are coupled~\cite{silsbee04}.
In particular the spin-Hall effect in two-dimensional electron systems, i.e.\
a spin-current which flows in the plane but perpendicular to the electrical current,
and which is polarized perpendicular to the plane, 
has been discussed intensively 
\cite{murakami03,sinova04,culcer03,sinitsyn03,shen03,schliemann04,burkov03,inoue04,murakami04,xiong04,nomura04,dimitrova04,chalaev04,mishchenko04,rashba04,khaetskii04}
over the last year.
The spin-Hall conductivity connects the spin-current 
with an electric field, $j_y^z = \sigma_{sH}E_x$, where $j_y^z$
denotes a current in $y$-direction with spin-polarization in $z$-direction.
In a clean two dimensional electron gas, the spin-Hall 
conductivity was predicted to have a universal value $\sigma_{sH} =  e/8 \pi$,
independent of the strength of the spin-orbit scattering \cite{sinova04}.
Several publications have addressed the issue of whether this result is modified in the
presence of impurity scattering.
Murakami \cite{murakami04} analyzed the 
Luttinger-Hamiltonian\cite{luttinger1956}, which applies to two-dimensional hole gases, and concluded
that the spin-Hall conductivity in the limit of weak impurity scattering reproduces
the intrinsic value (at least when restricting to s-wave impurity scattering).
For the Rashba model\cite{rashba1984}, which applies to two-dimensional electron gases, 
conflicting results exist in the literature:
By applying the standard Green's function techniques
Inoue et al.~\cite{inoue04} and Mishchenko et al.~\cite{mishchenko04} 
concluded that $s$-wave impurities suppress
the spin-Hall effect in bulk samples even when the disorder broadening of the energy levels is
small  compared to the spin-orbit splitting \cite{khaetskii04}.
On the other hand, Dimitrova \cite{dimitrova04} and Chalaev et al.~\cite{chalaev04},
starting from the same model Hamiltonian and applying similar methods, 
found a non-zero spin-Hall conductivity.
Even direct numerical evaluations of the effect do not  fully agree with each other:
Xiong and Xie \cite{xiong04} found within a scattering matrix approach the universal value of the
spin-Hall conductance $G_{sH}= e/8 \pi$ over a large parameter range.
Nomura et al.~\cite{nomura04} on the other hand found a spin-Hall conductivity of the order of 
but not identical to the universal value.

In this paper we calculate the spin-Hall conductivity for a bulk sample
within the self-consistent Born approximation.
We confirm Refs.~\cite{inoue04,burkov03,mishchenko04}, i.e., we find that\ even a weak disorder 
suppresses the spin-Hall conductivity.
For $s$-wave scatterers we calculate the impurity self-energy and
the dressed current vertex numerically. This allows us to 
obtain results beyond the limit $\epsilon_F \tau \to \infty$, which is accessible
analytically.
We find that a non-zero spin-Hall conductivity is, in principle, possible
although it remains much smaller than $e/8\pi$.

Our calculations are based on our previous work
\cite{raimondi2001,schwab2002}, where a number of technical details can be found. 
In the following we sketch the derivation of the spin-Hall conductivity.
The starting  point is the Hamiltonian
\begin{equation} \label{eq1}
H = \frac{p^2}{2m} + \alpha \mathbf{ \sigma } \cdot {\bf p} \times {\bf e}_z
,\end{equation}
where the parameter $\alpha$ describes the strength of the spin-orbit coupling,
$\sigma$ is a vector of Pauli matrices, and ${\bf e}_z$ is a unit vector perpendicular to the two dimensional 
system.
The spin-Hall conductivity is obtained by the standard linear response theory as
\begin{eqnarray}
\label{eq2}
  \sigma_{sH} & = &\lim_{\omega\rightarrow 0} \frac{e}{ \omega}
  \int \frac{ \D \epsilon }{2 \pi}
  \Tr \left[
  j_s^y \overline{ G^<(\epsilon) j_c^x G^A(\epsilon - \omega) }
  \right. \cr
&& \left. 
  + j_s^y \overline{ G^R(\epsilon) j_c^x G^<(\epsilon - \omega) }
  \right]
,\end{eqnarray}
with $G^<(\epsilon) = f(\epsilon) (G^R- G^A)$, $f(\epsilon)$ being the Fermi function.
In Eq.~(\ref{eq2}), the spin- and charge-current operators are given by 
   ${\bf j}_{s} = (1/4)\lbrace \sigma_z {\bf v} + {\bf v} \sigma_z \rbrace$ and
 ${\bf j}_{c} = {\bf v}$, respectively.     
The velocity operator, ${\bf v}$,  is obtained from the Hamiltonian (\ref{eq1}) and reads 
 $v^{x,y}=p^{x,y}/m\mp \alpha\sigma_{y,x}$. We choose the  electron charge as $-e$ $(e>0)$.
The trace in Eq.~(\ref{eq2}) is over the eigenstates of the Hamiltonian, and the 
bar indicates that the expression must be averaged over the disorder configurations.

When performing the  disorder average, we rely on the self-consistent Born approximation.
To begin with, we consider  point-like, i.e.,\ pure $s$-wave scatterers. 
The retarded/advanced impurity self-energy is then given by
\begin{equation} \label{eq3}
\Sigma^{R,A} = \frac{1}{2 \pi N_0  \tau}\sum_{\bf p}  G^{R,A}({\bf p})
.\end{equation}
Due to the spin-orbit coupling, the Green's functions have a non-trivial structure in the spin-space,
although the self-energy remains diagonal. Explicitly, one finds that
$\Sigma_{ss'}= \Sigma_0 \delta_{ss'}$,  $G_{ss'} = G_0 \delta_{ss'} + G_1 \sigma^x_{ss'} + G_2 \sigma^y_{ss'}$ with
\begin{eqnarray} 
\label{eq4}
G_0({\bf p}) & = & \frac{1}{2} \left(G_+ + G_- \right) \\
\label{eq5}
G_1( {\bf p})& = & \frac{1}{2} \frac{p_y}{p} \left(G_+ - G_-\right) \\
\label{eq6}
G_2( {\bf p})& = & - \frac{1}{2} \frac{p_x}{p}\left(G_+ - G_-\right) \\
\label{eq7}
G_{\pm}& = & \left( \epsilon + \mu - \frac{p^2}{2 m}  \mp \alpha p - \Sigma_0 \right)^{-1}.
\end{eqnarray}
By taking the  zero frequency limit of Eq.~(\ref{eq2}), the spin-Hall conductivity reads
\begin{equation} 
\label{eq8}
\sigma_{sH} = - \frac{e}{4\pi}
\sum_{\bf p}{\rm Tr}_\sigma \left[
2  j^y_s G^R({\bf p})J^x_c G^A({\bf p})\right]
,\end{equation}
since terms of the type $G^R G^R$ and $G^A G^A$ contribute only in the order
$(1/\epsilon_F \tau )(\alpha /v_F)^2 $ and can be safely neglected 
in the limit $\alpha p_F \ll \epsilon_F$ and/or  for weak disorder $\epsilon_F \tau \gg 1$.
The charge current $J^x_c$ has to be calculated including the vertex corrections,
$J^{x}_c=p^{x}/m+\Gamma^{x}$,
compare Eq.~(33) of Ref.~\cite{schwab2002}.
In the case of $s$-wave impurity scattering, the momentum dependent part of the current vertex
is not renormalized, while the momentum independent, but spin-dependent part,
 $\Gamma^{x}$, is obtained by solving the set of equations
\begin{equation} \label{eq9}
\Gamma_{ss'}^{x}=\gamma^{x}_{ss'}+\frac{1}{2\pi N_0\tau}\sum_{\bf p}\sum_{ab}
G^R_{sa}\Gamma_{ab}^{x}G^A_{bs'},
\end{equation}
with the {\sl effective} bare vertex given by 
\begin{equation} \label{eq10}
\gamma^{x}_{ss'}= - \alpha\sigma^{y}_{ss'}+\frac{1}{2 \pi N_0 \tau} \sum_{{\bf p},a}
G^R_{sa}({\bf p})\frac{p_x}{m} G^A_{as'}({\bf p}) .
\end{equation}
By expanding $\Gamma^x_{ss'}= \sum_\mu \Gamma^x_\mu \sigma^\mu_{ss'} $ in Pauli matrices
we obtain the spin-Hall conductivity as 
 \begin{equation} \label{eq11}
\sigma_{sH}=- \frac{e}{\pi}\, \Gamma^x_2 \,  \Im \sum_{\bf p }\frac{p_y}{m}G^R_0({\bf p}) G^A_1({\bf p}) \\
.\end{equation} 
Performing  the momentum integration Eq.~(\ref{eq11}) under the restriction that
$\alpha p_F \ll \epsilon_F$ and $\epsilon_F \tau \gg 1$
leads to  
\begin{equation} \label{eq11a}
  \sigma_{sH}= - \frac{e}{\pi}\, \Gamma^x_2 \,  \pi N_0 \tau \frac{\alpha p_F v_F \tau}{1+ 4 \alpha^2 p_F^2 \tau^2}
.\end{equation}
Apparently $\sigma_{sH}$ goes to zero when the spin-splitting, $\alpha p_F$, is small compared to the
disorder broadening of the levels, $1/\tau$.
If we neglect vertex corrections, i.e., if we insert in Eq.~(\ref{eq11}) 
the bare vertex $\Gamma^x_2=-\alpha$, we find
\begin{equation} \label{eqBare}
  \sigma_{sH}\big|_{\rm bare \, vertex}=
  \frac{e}{8\pi} \frac{4 \alpha^2 p_F^2 \tau^2}{1+4\alpha^2p_F^2\tau^2}
 \end{equation} 
i.e., the universal value  $\sigma_{sH} =e/8\pi$ is recovered
in the weak disorder limit. On the other hand, by inserting the dressed vertex   $\Gamma_2^x \approx 0$ as
calculated in Refs.~\cite{raimondi2001,schwab2002} one finds that $\sigma_{sH} \approx 0$. 
Our result then agrees with that found in \cite{inoue04,burkov03,mishchenko04}.

\begin{figure}
  \centerline{\includegraphics[height=5.0cm]{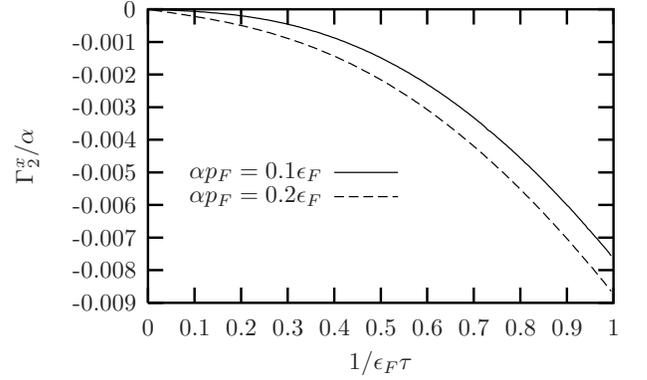}}
  \caption{\label{fig1} 
  The dressed vertex $\Gamma^x_2$ in units of $\alpha$ as function of
  disorder strength $1/(\epsilon_F\tau )$. 
  $\Gamma_2^x$ enters the dressed charge current, 
  $J_c^x = p_x/m  +  \Gamma_2^x \sigma_y $ and thus the spin-Hall conductivity, Eq.~(\ref{eq11a}).
  In comparison with the bare chare current, $j_c^x = p_x/m  -\alpha \sigma_y $, 
  the spin-dependence is strongly reduced. }
\end{figure}

As explained in Ref.~\cite{schwab2002}, the vanishing of the
dressed vertex $\Gamma^x_2$ is due to the fact that 
the integral on the right-hand side of Eq.~(\ref{eq10}) gives 
$\approx\alpha \sigma^y$,
making the {\sl effective} bare vertex $\gamma$ itself to vanish.
A more careful numerical evaluation of the integral for arbitrary
disorder strength actually shows that the compensation
of the two terms in Eq.~(\ref{eq10}) is exact only in the weak disorder limit,
i.e., \ $\epsilon_F \tau \gg 1$. In Fig.~\ref{fig1} we show the
dressed vertex as a function of disorder.
$\Gamma_2^x$ goes to zero as $1/\epsilon_F \tau \to 0 $, nothing
special is observed as $\alpha p_F \sim 1/\tau$, and
even in the strong disorder limit
$\Gamma_2^x$ remains much smaller than its bare value ($-\alpha$). 
We conclude that although, in principle, a non-zero spin-Hall
conductivity may be obtained,  one expects a much
smaller value than the universal one. 

Next we address the question whether the spin-Hall effect is sensitive to the type of disorder potential.
Inoue et al.\cite{inoue04} argued that $\sigma_{sH}$ may be non-zero for long-range defect potentials,
although an explicit result has not been given.
In our calculation we follow again
closely \cite{schwab2002} where all the details can be found.
Here we assume weak disorder, so that the inequalities $\epsilon_F \gg \alpha p_F \gg 1/\tau$ hold.
In the following, we work in the  eigenstate basis of the Hamiltonian (\ref{eq1}) 
\begin{equation}
| {\bf p} \pm \rangle = \frac{1}{\sqrt{2}} \left\{ \pm {\rm i}
\exp(-{\rm i }\varphi )   | {\bf p } \uparrow \rangle  +  |{\bf p} \downarrow \rangle       \right\}
\end{equation} 
where $\tan(\varphi)= p_y/p_x $ and the corresponding eigenvalues are $E_{\pm} = p^2/2m \pm \alpha p$.
In this basis, the matrix elements of the current operators read 
\begin{eqnarray}
   \langle {\bf p } \pm   | j^s_y |{\bf p} \pm  \rangle &=& 0  \\
   \langle {\bf p } \pm   | j^s_y |{\bf p} \mp  \rangle &=&  -\frac{1}{2}\frac{p}{m} \sin(  \varphi ) \\
   \langle {\bf p } \pm  |  j^x_c |{\bf p} \pm  \rangle &=&  \left( \frac{p}{m} \pm \alpha \right) \cos(\varphi ) \\
   \langle {\bf p } \mp  | j^x_c |{\bf p} \pm   \rangle &=&  \mp \I \alpha \sin(\varphi) 
.\end{eqnarray}
To use Eq.~(\ref{eq8}) we need the dressed charge operator, $J^x_c$.
Since, as seen from the above equations, the spin-current operator is off-diagonal 
in the eigenstate basis
we get the spin-Hall conductivity in the form
\begin{equation}
  \sigma_{sH} = - \frac{e}{\pi}
  \sum_{\bf p} \Re \left[
  \langle {\bf p } + | j^y_s |{\bf p} - \rangle 
  \langle {\bf p } - | J^x_c |{\bf p} + \rangle
 G^R_-({\bf p}) 
 G^A_+({\bf p}) \right]
.\end{equation}
To calculate the dressed current operator we make use of the assumption that the spin-orbit splitting
is large compared to the impurity broadening of the levels, $\alpha p_F \gg 1/\tau$. 
The off-diagonal matrix elements of the current operator are then obtained in terms of
the diagonal ones,
\begin{eqnarray}
\lefteqn{ \langle {\bf p }  \mp  | J^x_c |{\bf p} \pm    \rangle  =
\langle {\bf p }  \mp  | j^x_c |{\bf p}  \pm   \rangle}  \nonumber\\[0.5mm]
&&
+ \sum_{{\bf p'}, m} \Big[  \langle {\bf p }  \mp  | V |{\bf p}'  m  \rangle  \,
                    \langle {\bf p}'  m  | V |{\bf p}   \pm   \rangle  \nonumber\\[0.5mm]
&& \times G^R_{m}({\bf p}')G^A_m({\bf p}') \langle {\bf p' } m  | J^x_c |{\bf p}'  m   \rangle  \Big]
.\end{eqnarray}
The diagonal matrix elements on the other hand 
were already considered in Ref.~\cite{schwab2002}, 
and are obtained from the equation
\begin{eqnarray}
\lefteqn{ \langle {\bf p }  \pm  | J^x_c |{\bf p}  \pm   \rangle  = 
\langle {\bf p }  \pm  | j^x_c |{\bf p}  \pm  \rangle }  \\
&&
+ \sum_{{\bf p'}, m}   |\langle {\bf p }  \pm  | V |{\bf p}'  m  \rangle |^2 \,
G^R_{m}({\bf p}')G^A_m ({\bf p}') \langle {\bf p' } m  | J^x_c |{\bf p}'  m   \rangle  \nonumber
.\end{eqnarray}
We consider impurity scattering which conserves spin, but allow the scattering
amplitude to be  momentum-transfer dependent. 
Such a dependence appears as a product of two contributions.
The first is due to the type of disorder potential one considers,  $ V_{{\bf p}, {\bf p}'}$,
while the second is induced by the transformation to the eigenstate basis.
The latter gives rise to the following matrix elements
\begin{eqnarray}
\langle {\bf p }  -  | V |{\bf p}'  \pm  \rangle  \,
                    \langle {\bf p}'  \pm   | V |{\bf p}   +   \rangle  = 
 \mp \frac{\I}{2} \sin(\varphi-\varphi' ) | V_{{\bf p}, {\bf p}'} |^2  
\end{eqnarray}
and
\begin{eqnarray}
| \langle {\bf p }  \pm | V |{\bf p}'  \pm  \rangle |^2  &= &  \frac{1}{2} | V_{{\bf p}, {\bf p}'} |^2 
(1+\cos( \varphi-\varphi' ) ) \\
| \langle {\bf p }  \pm | V |{\bf p}'  \mp   \rangle |^2 & = & \frac{1}{2} | V_{{\bf p}, {\bf p}'} |^2
(1-\cos( \varphi-\varphi' ) ) 
\end{eqnarray}
We assume that the scattering amplitude $ V_{{\bf p}, {\bf p}'}$
depends weakly on the momentum transfer, so 
that the scattering probability depends only on the angle between the incoming and scattered particle.
Under this condition we can expand the scattering probability as
\begin{equation}
  | V_{ {\bf p}, {\bf p}'} |^2 = V_0 + 2 V_1  \cos(\varphi - \varphi') + 2 V_2 \cos( 2 \varphi - 2 \varphi' ) + \dots
.\end{equation}
To obtain the spin-Hall conductivity the two momentum integrations over ${\bf p}$ and
${\bf p}'$ have to be performed.
We split the momentum integration in an integral over the energy $\xi = p^2/2m -\mu$ and
the angular variable $\varphi$,
\begin{equation}
   \sum_{\bf p} \to N_0 \int \D \xi \int \frac{\D \varphi}{ 2 \pi}
,\end{equation}
and find
\begin{eqnarray}
N_0 \int \D \xi G^R_- G^A_+ &= & \frac{2 \pi \I N_0 }{2 \alpha p_F +\I/\tau} \approx \frac{\I \pi N_0 }{ \alpha p_F} \\
N_0 \int \D \xi G^R_{\pm} G^A_{\pm} &=  &2 \pi N_\pm \tau_\pm
,\end{eqnarray}
where $N_\pm$ and $\tau_\pm$
are the density of states and the lifetime in the two subbands. 
Notice that the first of the two integrals is correct only to the lowest order in the (small) parameter
$\alpha p_F /\epsilon_F$, whereas the second integration is valid beyond that limit. 
Finally the spin-Hall conductivity is determined as
\begin{eqnarray} \label{eq29}
\sigma_{sH} = \frac{e}{8\pi}\left[ 1 - \frac{1 }{4 \alpha}
             \frac{N_+ \tau_+ J^x_+ - N_- \tau_- J^x_-}{N_0 \tau}\frac{ V_2 - V_0 }{V_0} \right] 
,\end{eqnarray}
where the second term is due to vertex corrections. When calculating  the product
$N_\pm \tau_\pm J^x_\pm $ to zero order in $\alpha p_F/\epsilon_F$ 
the vertex corrections disappear.
Expanding the density of states, the scattering time and to dressed current operator 
to first order yields
\begin{eqnarray} 
N_\pm & \approx &    N_0 \left( 1 \mp  \frac{\alpha p_F }{ 2 \epsilon_F} \right) \\
\tau_\pm & \approx &\tau \left( 1 \pm  \frac{V_1}{V_0}\frac{\alpha p_F}{2 \epsilon_F }  \right)  \\
J^x_\pm &\approx & \frac{V_0}{V_0 -V_1} \frac{p_F}{m} \mp \alpha \frac{V_0 + V_2}{V_0 -V_2}. 
\end{eqnarray}
Notice that the dressed current operator, to  the leading order in $\alpha$, is 
of the familiar form $J = j  \tau_{\rm tr}/\tau$, where $\tau_{\rm tr}$ is
the transport scattering time.
By  combining all the terms, one then finds that the spin-Hall conductivity (\ref{eq29})
vanishes as in the case of pure $s$-wave scattering, 
$\sigma_{sH} = 0$.

As a  last useful observation,  we separate the reactive and dissipative
contributions to the current response, $\sigma_{sH}= \sigma_{sH}^R + \sigma_{sH}^D$ where
\begin{eqnarray}
  \sigma_{sH}^R & = & \lim_{\omega\rightarrow 0}\frac{e}{ \omega}
  \int \frac{ \D \epsilon }{2 \pi}
  \Tr \left[
  j_s^y \overline{ G^<(\epsilon) j_c^x  \Re G^A(\epsilon - \omega) }
  \right.  \nonumber \\[1mm]
&& \left.
  + j_s^y \overline{\Re G^R(\epsilon) j_c^x G^<(\epsilon - \omega) }
  \right]
  \\[1mm]
  \sigma_{sH}^D & = & - \frac{e}{\pi}
  {\rm Tr}\left[
  \overline{  j^y_s \Im G^R j^x_c \Im G^R } \right]
.\end{eqnarray}
Since the zero frequency spin-Hall conductivity is real, 
the terms with imaginary (real)  current matrix elements  
 contribute to  $\sigma^R_{sH}$ ($\sigma^D_{sH}$), respectively.
It then follows  that the first term on the right hand side of Eq.~(\ref{eq29})
corresponds to $\sigma^R_{sH}= e/ 8 \pi$, whereas
the second term (the vertex corrections) is the dissipative response with
\begin{equation}
\sigma_{sH}^D = -\frac{e}{8\pi}\frac{1 }{4 \alpha}
             \frac{N_+ \tau_+ J^x_+ - N_- \tau_- J^x_-}{N_0 \tau}\frac{ V_2 - V_0 }{V_0} 
 = -\frac{e}{8\pi}
\end{equation}
and only the sum of the reactive and dissipative response is zero.

In summary, 
we calculated the spin-Hall conductivity in a two dimensional electron gas
within the self-consistent Born approximation, including the vertex corrections in the ladder 
approximation.
We remark that, although a
number of similar studies exist in the recent literature, the final conclusions
are often contradictory.  This may be due to the fact that 
the relevant integrals depend in a very subtle way on the type
of the physical limit considered. For this reason in this work we evaluated 
all the relevant integrals both
analytically and numerically. This allowed us to confirm
the conclusions of Refs.~\cite{inoue04,mishchenko04,burkov03}.
In particular, we find that the spin-Hall conductivity is strongly suppressed below the
universal value of $e/8\pi$.
Furthermore we have demonstrated that the result is not only valid for pure $s$-wave scattering, but is
robust upon the inclusion of  a weak momentum dependence of the scattering probability.


\begin{thebibliography}{99} 
\bibitem{silsbee04}R.~H.\ Silsbee, J.\ Phys.: Condens.\ Matter {\bf 16} R179 (2004).
\bibitem{murakami03}S.\ Murakami, N.\ Nagaosa, and S.-C.\ Zhang, Science {\bf 301}, 1348 (2003);
 Phys.\ Rev. B {\bf 69} 235206 (2004).
\bibitem{sinova04}J.\ Sinova, D.\ Culcer, Q.\ Niu, N. \ A. \ Sinitsyn,
T. \ Jungwirth, and A.\ H. \ MacDonald,  Phys.\ Rev.\ Lett.\ {\bf 92}, 126603 (2004).
\bibitem{culcer03} D.\ Culcer, J.\ Sinova, N.~A.\ Sinitsyn, 
T.\ Jungwirth, A.~H.\ MacDonald, Q.\ Niu, Phys.\ Rev.\ Lett.\ {\bf 93}, 046602 (2004).
\bibitem{sinitsyn03}N.~A.\ Sinitsyn, E.~M.\ Hankiewicz, W.\ Teizer, J.\ Sinova,
    cond-mat/0310315 (unpublished).
\bibitem{shen03}S.\ Shen, cond-mat/0310368 (unpublished); 
L.\ Hu, J.\ Gao, and S.\ Shen, cond-mat/0401231 (unpublished).
\bibitem{schliemann04}J.\ Schliemann and D.\ Loss, Phys.\ Rev.\ B {\bf 69}, 165315 (2004);
    cond-mat/0405436 (unpublished). 
\bibitem{burkov03}A.~A.\ Burkov, A.~S.\ Nu{n}ez and A.~H.\ MacDonald, cond-mat/0311328 (unpublished).
\bibitem{xiong04}Ye Xiong and X.~C.\ Xie, cond-mat/0403083 (unpublished).
\bibitem{nomura04}K.\ Nomura et al., cond-mat/0407279 (unpublished).
\bibitem{inoue04}J.\ Inoue, G.~E.~W.\ Bauer, and L.~W.\ Molenkamp, 
    Phys.\ Rev.\ B {\bf 70}, 041303(R) (2004).
\bibitem{murakami04}S.\ Murakami, Phys.\ Rev.\ B {\bf 69}, 241202(R) (2004).
\bibitem{dimitrova04}O.~V.\ Dimitrova, cond-mat/0405339 (unpublished).
\bibitem{chalaev04}O.\ Chalaev and D.\ Loss, cond-mat/0407342 (unpublished).
\bibitem{mishchenko04}R.~G.\ Mishchenko, A.~V.\ Shytov, and B.~I.\ Halperin, cond-mat/0406730 (unpublished).
\bibitem{rashba04}E.~I.\ Rashba, cond-mat/0404723 (unpublished).
\bibitem{khaetskii04}Most recently, also A.\ Khaetskii, cond-mat/0408136 (unpublished), confirmed this result.
\bibitem{luttinger1956} J.~M.~Luttinger, Phys.\ Rev.\  {\bf 102}, 1030 (1956).
\bibitem{rashba1984}Y.~A.~Bychkov and E.~I.~Rashba, J.\ Phys.\ C {\bf 17}, 6039 (1984).
\bibitem{raimondi2001}R.\ Raimondi, M.\ Leadbeater,  P.\ Schwab, E.\ Caroti, and C.\ Castellani,
Phys.~Rev.~B {\bf 64}, 235110 (2001).
\bibitem{schwab2002}P.\ Schwab and R.\ Raimondi, Eur.~Phys.~J.\ B {\bf 25}, 483 (2002).
\end{thebibliography}
\end{document}